%% file: WAD.tex
\documentclass[mypaper,7pt,twoside]{CoAst}
\usepackage{epsf,graphicx,fancyhdr}
\usepackage[centertags]{amsmath}
\usepackage{amsfonts}
\usepackage{amssymb}
\usepackage{amsthm}
\input{CoAst_liege-proceedingslayo}

\begin{document}
\sf

\chapterCoAst{
Driving mechanism in massive B-type pulsators }{W.A. Dziembowski}

\Authors{W.A. Dziembowski} \Address{Warsaw University Observatory,
Al. Ujazdowskie 4, 00-478 and \\Copernicus Astronomical Center,
Bartycka 18, 00-716 Warsaw, Poland}

\noindent
\begin{abstract}
After a historical introduction, I present the current status of our
understanding of the mechanism responsible for pulsation
in $\beta$ Cephei and SPB stars.

\end{abstract}
\section{Introduction}
Variability of the certain $\beta$ Cephei  stars has been known since the beginning of previous
century but serious efforts to explain its origin were undertaken much later.
To my knowledge Paul Ledoux was the first theorist who got interested in these objects
and the first who invoked  nonradial oscillations to explain pulsation in any stars.
In his pioneering work, Ledoux (1951) used data on spectral line profile changes
too reveal nature of modes responsible for variability of prototype
star with two close frequencies. More than ten years later, Chandrasekhar
\& Lebovitz (1962) proposed a different interpretation of the two frequencies.
These two papers had a lasting impact on studies of nonradial oscillations in stars.
However, the matter of driving was only briefly touched upon.
The search for the driving mechanism began few years after.
There has been a number of proposals suggesting
the deep interior as the site of the driving effect. As it turned out, this
was not correct.

The road toward correct answer was initiated by Stellingwerf (1978).
He noted that a slight bump in opacity connected with the HeII ionization edge at $T=1.5\times 10^5$K produces a
substantial driving effect in $\beta$ Cep models. The effect was not
big enough to cause an instability of any mode in any of the models.
However, Stellingwerf suggested that an improvement in
the opacity calculations might lead to an enhancement of the bump and consequently
to instability of modes corresponding to $\beta$ Cep pulsations.
Subsequently, Simon (1982) pointed out that augmenting the heavy element opacities
by the 2--3 factor would resolve the period ratio discrepancy in
classical Cepheid, which has been another outstanding puzzle in stellar
pulsation theory, in addition  to  significant enhancement of the bump postulated by Stellingwerf.
Simon's plea for reexamination of the heavy element
contribution to stellar opacity has been an inspiration for the OPAL (Iglesias {\it et
al.} 1987) and the OP (Seaton 1993) projects. Inclusion of hitherto neglected transitions in heavy
element ions resulted in large (up to factor 3) increase of opacities in the temperature range
near $T\approx2\times 10^5$K leading to a pronounced bump, which is often referred to as the the Fe-bump,
because the transitions within the iron M-shell are the primary contributors.

Not long after the OPAL opacities became available for stellar
modeling, first papers demonstrating that there are unstable modes
in $\beta$ Cep stars with periods consistent with observations were
published (Cox {\it et al.} 1992; Kiriakidis {\it et al.} 1992;
Moskalik \& Dziembowski 1992). However, as pointed in the last paper, explanation of
pulsation in lower luminosity $\beta$ Cep required metallicity parameter $Z\geq0.03$, which
 appeared too large even then.
An improvement in atomic physics introduced in the subsequent release of the OPAL opacities
(Rogers \& Iglesias, 1992) removed this discrepancy. In the first models employing
the new opacity data (Dziembowski \& Pamyatnykh 1993;
Gautschy \& Saio 1993; Dziembowski {\it et al.} 1993) the instability was found already
at $Z=0.02$.
Moreover, in addition to low-order p- and g-mode instability
responsible for $\beta$ Cep pulsation, an instability of high-order
g-modes in a detached lower frequency range was found. In the the $\beta$
Cep domains instability was found only at high angular degrees
($\ell\ge6$). However, for lower luminosity stars the instability
extended to more easily detectable low degree modes and this
provided a natural explanation of the origin of pulsation in Slowly
Pulsating B (SPB) stars. This new type of variable stars was defined
only two years earlier by Waelkens (1991).

\section{Opacity mechanism in B stars}

\figureCoAst{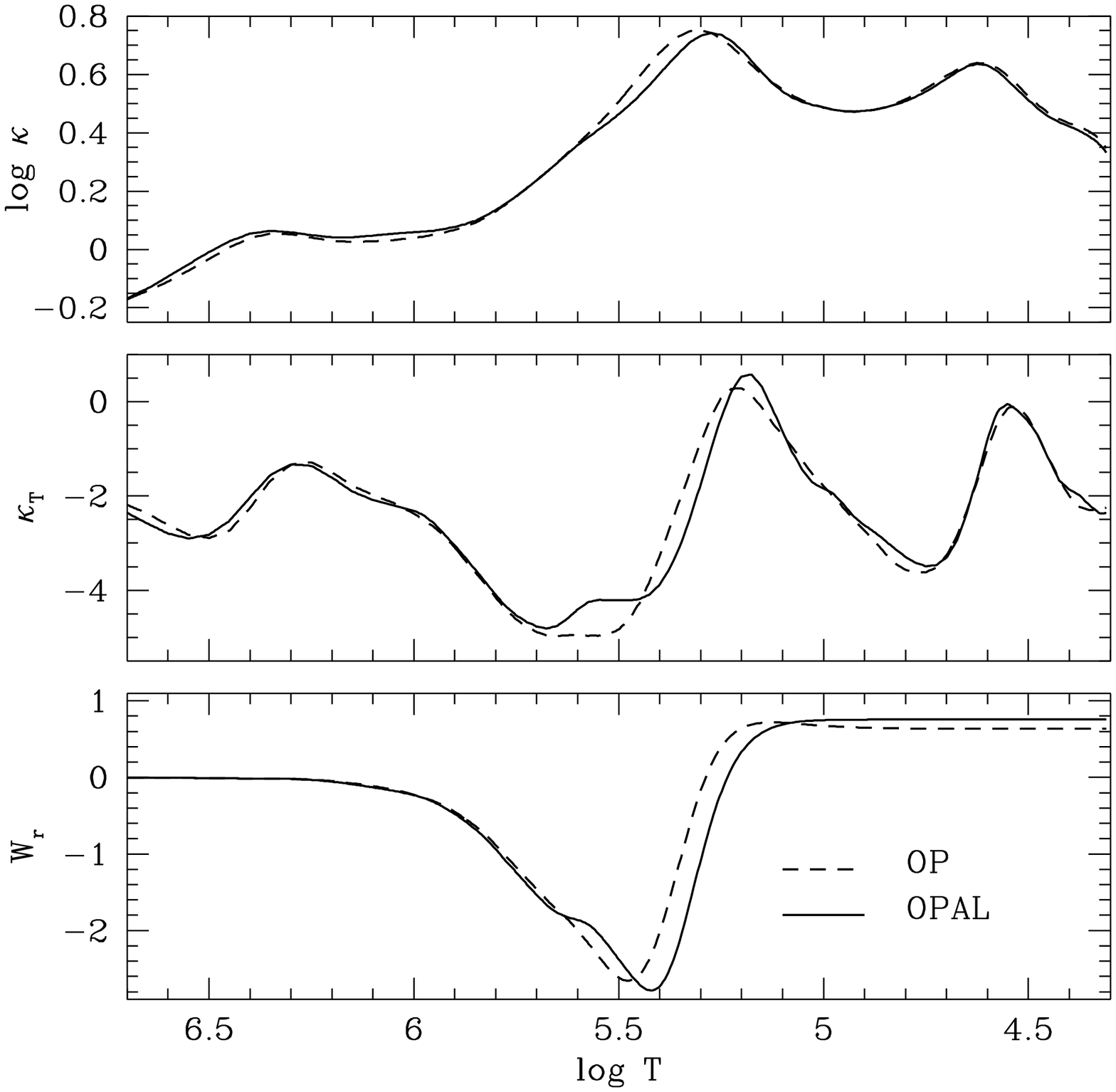}{ Panels from top to bottom show Rosseland mean
opacity, $\kappa$, its logarithmic temperature derivative, $\kappa_T$, and the
cumulative work integral for the fundamental radial mode, $W_r$,
plotted against temperature in two stellar models. The models are
characterized by the same parameters: $M=9.63M_\odot$, $\log T_{\rm
eff}=4.336$, $\log L=3.891$ $X=0.7$, $Z=0.0185$, and the same heavy
element mix (Asplund et al. 2005), but use different opacity data.
Results obtained with the OPAL data (Iglesias \& Rogers 1996) are
shown with solid lines and those and those obtained with the OP data
(Seaton 2005) are shown with dashed lines. }
{W}{!ht}{clip,angle=0,width=100mm}

Not much was added after 1993 to our basic understanding how the
opacity mechanism in massive B stars works. Plots which I show here
in Figure 1 are very similar to those shown in Fig.1 of
Dziembowski \& Pamyatnykh (1993). The two envelope models selected
for the plots correspond to an object in the mid of the $\beta$ Cep
domain in the H-R diagram and they differ only in the opacity data.
Here, the newest versions of the data from the OPAL and OP projects
were adopted. Shown in the upper panel, the run of the opacity
coefficient, $\kappa$, reveals differences between the two data.  The
largest difference is seen at the inner side of the Fe-bump, which is
centered at $\log T\approx5.3$. Centered at $\log T\approx4.6$, the
bump caused by the HeII ionization plays no role in driving B star pulsation.
The bump at $\log T\approx6.4$, which is associated mainly with
L-shell transitions and ultimate ionization of C, O, and Ne
may be active in still hotter stars.

To understand how pulsation is driven, more helpful than the plot of opacity itself
is the plot of its logarithmic temperature derivative at constant density,
$\kappa_T$, which is depicted in the mid plot. The close connection
of $\kappa_T$ with mode excitation is seen when its run is
compared with  the run of the normalized cumulative work
integral, $W_r$, which is shown in the bottom panel. The latter
describes the pulsation energy gain or loss by a mode per unit of
time between the center and the distance $r$. An expression for $W_r$ in terms of
eigenfunction describing the Lagrangian perturbation of temperature
and total flux, respectively, $\delta T$ and $\delta L$, may be
written as follows,
$$
W_r=-{1\over L}\int_0^r{\rm d}r\oint{\rm d}t \Re\left[\left({\delta T\over
T}\right)^* {{\rm d}\delta L\over{\rm d}r}\right] =\int_0^r{\rm
d}r\left\vert{\delta T\over T}\right\vert^2{{\rm d}\kappa_T\over{\rm
d}r}+....$$
where in the second equality, I wrote explicitly only the term
giving rise to the opacity effect. This is just one of several terms
arising from $\delta L$ but it is the one that matters here. We
may see in Figure 1 that in the driving zone, where $W_r$ increases,
$\kappa_T$ increases too and the opposite is true in the damping zone.
It is the slope of $\kappa_T$ and not its value which is really
relevant. In our models there are three driving slopes, but the only
one active occurs in the thin layer extending $\log T\approx5.5$ to
5.2 and is associated with the Fe-bump. The remaining two are
inactive for different reasons. In the layer of the deep bump,
the pulsation amplitude is very low while in the HeII-bump zone
the thermal relaxation time, $\tau_r$, is much shorter than the pulsation
period, $\Pi$, so that the $\delta L$ gradient cannot be maintained. All the
damping arises in the layer of decreasing $\kappa_T$ below
the Fe-bump. In the considered cases, it is overcompensated by the driving
above which renders the mode unstable.

The conditions for the mode instability are the same as for all
opacity-driven pulsation. Within the zone of rising $\kappa_T$, the mode
amplitude must be large and slowly varying with distance, so that
the opacity perturbation dominates in the perturbed flux, and the
thermal relaxation time is not significantly shorter than the
pulsation period ($\tau_r\gtrsim\Pi$). Outside such a zone, the amplitude
must be low or we must have $\tau_r<<\Pi$. In B stars we
encounter the opacity mechanism in the cleanest form. There is a
convective layer in the Fe-bump zone of the early B-type stars but
by far most of the flux is carried by radiation. The adiabatic
temperature gradient changes within this zone but the variations are
small, thus there is no significant role of the $\gamma$-mechanism.

In the two models considered, only one radial mode is unstable. At
somewhat higher effective temperature, the
first overtone is unstable, in addition to the fundamental.
In our two models, the number of unstable nonradial modes is huge and the same is
true for nearly all models of SPB and $\beta$ Cep stars.
The change in mode geometry does not change conditions for instability. The occurrence of the two
instability ranges may be easily understood by considering changes
in the shape of the radial eigenfunction describing Lagrangian
perturbation of  pressure, $y_p(r)\propto\delta p/p$, with period,
$\Pi$, at specified degree, $\ell$. For greater generality, it is
better to consider changes with the dimensionless frequency,
$$\sigma\equiv{2\pi\over \Pi}\sqrt{R^3\over GM}.$$

 \figureCoAst{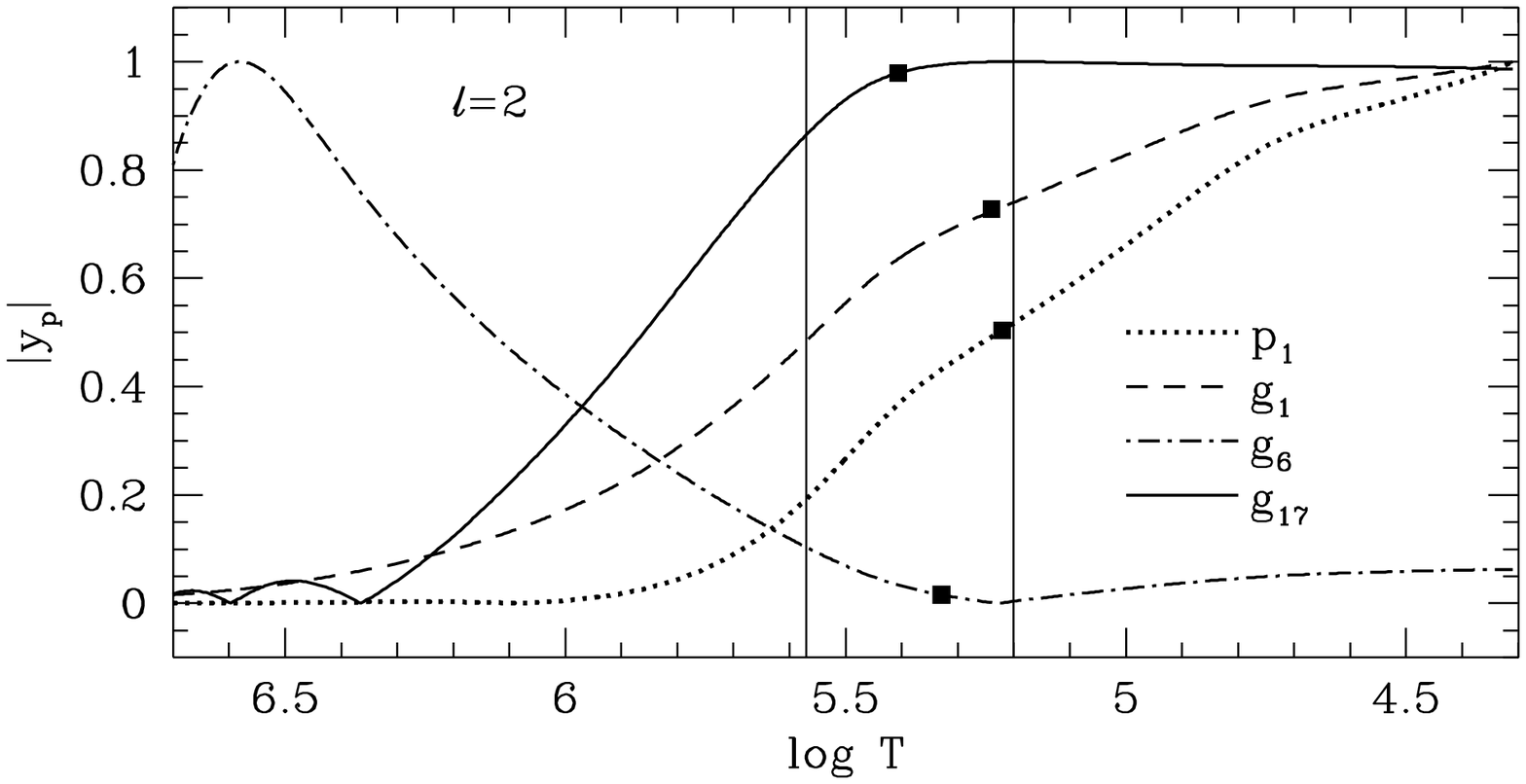}{Absolute value of the eigenfunction
describing Lagrangian pressure perturbation for selected quadrupole
modes in the model calculated with the OP opacity data. The model
parameters are given in the caption of Figure 1. Modes p$_1$ ($\Pi=0.115$d)
and g$_6$ (0.529 d) are stable while modes g$_1$ (0.175 d) and g$_{17}$ (1.376 d) are unstable. Dots in the
curves indicate places where $\tau_r=\Pi$. The two vertical lines
mark the boundaries of the driving slope associated with the
Fe-bump.} {l2}{!ht}{clip,angle=0,width=100mm}

Let us focus on the  model employing the OP data. This is one
of the seismic models considered  for the $\beta$ Cep
star $\nu$ Eri (Dziembowski \& Pamyatnykh 2008). Behavior of $y_p$
in the outer layers for four selected quadrupole modes is shown in
Figure 2. Mode $p_1$ has the period similar to first radial overtone
and, like that, is stable. As long as $\ell\ll\sigma^2$, which is true in
this case, the $\ell$ value has little effect on $y_p$ in the outer
layers. Mode g$_1$ has the shape of $y_p(r)$ and period
suitable for driving. Unstable are also g$_2$ and g$_3$
modes. Periods of these three modes are in the range of 0.14 to
0.28 d, which is typical for $\beta$ Cep stars. At $\ell=1$ and
 ($3\le\ell\le8$) there are two or three
unstable modes. Then the number decreases to 1 and at $\ell=11$
the instability disappears.

In the intermediate frequency range ($\sigma\sim\sqrt{\ell}$), the
absolute maximum of $|y_p|$ occurs below the Fe-bump. The mode g$_6$
in Figure 2 is an example. This mode is stable, though there is a
significant destabilizing contribution arising at the driving slope
of the deep bump but it compensates less than 60\% of damping
occurring below. The layers above $\log T=6.3$ bring virtually no
contribution to the total work integral. The $\ell=2$ g-modes
between $n=5$ and $n=13$ are stable. At still lower frequencies, the
maximum is again close to the surface and a new instability range
may appear. All the $\ell=2$ modes in the range $13\le n\le20$ are
unstable. We may see in Figure 2 that the g$_{17}$ mode has the
shape $y_p$ perfect for driving. Similar shapes are found for other
degrees if $0.06\lesssim\sigma^2/\ell\lesssim0.3$, which implies
shorter periods and higher radial orders at higher degrees. In our
model, there are no unstable modes of this type at $\ell=1$ because
the $\tau_r\gtrsim\Pi$ condition is not well satisfied along the
driving slope. The number of unstable modes increases up to
$\ell=24$ and the instability disappears at $\ell=35$. With the
$m$-dependence included, there is about 28 000 unstable high-order
g-modes, which is nearly two orders of magnitude more than the
low-order modes in this $\beta$ Cep star model.

In the model calculated with the OPAL opacity data, the slow mode
instability begins only at $\ell=4$. This shift is connected with
the upward shift of the Fe-bump (see Figure 1), hence lower $\tau_r$
within its range. Modes of lower $\ell$s with suitable $y_p(r)$ have
too long period to satisfy $\tau_r\gtrsim\Pi$ condition. To find
instability with the OPAL data, we have to go to stars with lower
$L$ and/or $T_{\rm eff}$ where the Fe-bump occurs at higher density,
hence at higher $\tau_r$. This explains, as Pamyatnykh (1999) first
noted, a significant difference between blue boundaries of the
instability domains of low degree modes in the H-R diagrams
calculated with OPAL and OP data. The latter place the boundaries,
both of SPB and $\beta$ Cep at a higher $T_{\rm eff}$. Miglio et al.
(2007), who used the new OP data (Seaton 2005) found difference ln
$\log T_{\rm eff}$ reaching up to 0.05. The two domains partially
overlap.

\section{Question that remain to be answered}

\subsection{Are the current opacity calculations adequate?}

The theory should account for excitation of all detected modes in
the star and still it does not in all the cases. Particularly
challenging are hybrid objects, where both high and low frequency
modes are found. One such object is $\nu$ Eri. The difficulty with
explaining excitation of modes at both ends of the observed
frequency range were discussed by Pamyatnykh et al. (2004), who
suggested that the iron abundance in the driving layer is
significantly enhanced due to selective radiation pressure. The
enhancement required to destabilize high frequency modes was
somewhat less than factor 4 and somewhat larger for the low
frequency mode. The authors based their proposal on the Charpinet et
al. (1996) solution of driving problem for sdB pulsators.
Unfortunately, they ignored work of Seaton (1999) from which it
clearly follows that in massive B stars levitation leads to
enhancement of the iron abundance not only in the bump zone but also
in photosphere, and thus cannot be hidden.

We revisited the problem of mode excitation in $\nu$ Eri and in
another hybrid pulsator, 12 Lac, in our recent paper (Dziembowski \&
Pamyatnykh 2008). Since there is no spectroscopic evidence for chemical
anomalies in any of these objects, we argued that levitation
must be offset by a macroscopic mixing. We showed that even in the very
slowly rotating $\nu$ Eri, mixing in outer layers by meridional
circulation may be fast enough. With the OP opacities, there are
unstable high- order g-modes $\ell=2$ and their periods are in the
1.1 -1.6 d range in the $\nu$ Eri model. The observed periods are
1.6 and 2.3d. This might suggest that it is only a matter of
further improvement in the opacity calculations to get the
agreement. This is possible. However, in the case of 12 Lac, which
is a brighter object, explanation of the long period (2.8 d),
requires much larger opacity modification. Moreover, the use of the
OP data did not help a bit in solving the difficulty with mode
driving at the short period end. These discrepancies may justify a
new plea to atomic physicists for revisiting opacity calculations.

\subsection{How rotation affects driving}

The effect of Coriolis force becomes significant as soon as the spin
parameter, $s\equiv2\Pi/\Pi_{\rm rot}$, approaches one and, for the
high-order g-modes, this happens well before the rotation rate
approaches the maximum value. Modes cannot be described in spherical
harmonics but within the {\it traditional approximation}, separation
of the radial and angular dependencies in the pulsation amplitudes
is still possible in terms of the {\it Hough functions}. Then, the
Coriolis force effect is reduced to the replacement
$\ell(\ell+1)\rightarrow\lambda(s)$. The essence of the driving
effect is not changed. The range of models having unstable low
degree modes is somewhat increased (Townsend 2005). However, this
approximation may be inadequate (Lee \& Saio  1989). There are
intriguing discoveries of a large number of modes in four Be stars
with the {\it MOST} satellite and discrepant conclusions from model
calculations regarding stability of these mode. (Cameron et al.
2008, and references therein). Accurate modeling of oscillations in
these extreme rotators is still ahead of us. This is important
because the detected modes may yield us the clue to understanding
the Be star phenomenon.

\subsection{What is the role of high-degree modes }

A vast majority of the unstable modes are of high degree $\ell>4$ and, thus, cannot be easily detected
if their intrinsic pulsation amplitudes are similar to those of those low degrees.
Smolec and Moskalik (2007) proposed that such modes play a role in collective
saturation of the driving. This was their explanation why the amplitude they determined by nonlinear
modeling of radial pulsation in $\beta$ Cep stars  were much higher than observed in any star of this type.
They noted, however, that the postulated high-$\ell$ modes may be difficult to hide, as they should
contribute to spectral line broadening. According to my crude estimate, if saturation is mostly
due to excitation of g-modes of high orders and degrees, then the r.m.s. velocity in the atmosphere should be
between 50 and 100 km/s. This seems unacceptably high and, thus, we must conclude that the instability is not saturated.
In such a case the terminal state of pulsation must be determined  by a resonant excitation of damped modes.
If all unstable modes had the same chances to grow, then the occurrence of detectable pulsation would have be
regarded a miracle. The modes with $\ell\le 2$ constitute only about 0.2 percent of the total population.
Apparently, the coupling is more effective for high degree modes. Questions why it is so and what is the contribution
of the invisible modes to spectral line broadening are awaiting answers.

\acknowledgments{
This paper was supported by
the Polish MNiSW Grant No.~1~P03D~011~28.}

\References{
Asplund M., Grevesse N., Sauval A.J. 2005, ASP Conf. Ser., Vol. 336, p. 25\\
Cameron C., Saio H., Kuschnig R. et al. 2008, arXiv:0805.1720 (astro-ph)\\
Chandrasekhar, S., Lebovitz, N. 1996, ApJ 136, 1105\\
Charpinet S., Fontaine G., Brassard P., Dorman B., 1996, ApJ, 471, L103\\
Cox, A.~N., Morgan, S. M., Rogers, F. J., Iglesias, C.~A. 1992, ApJ 393, 272\\
Dziembowski, W. A., Pamyatnykh, A. A. 1993, MNRAS 262, 204\\
Dziembowski, W. A., Pamyatnykh, A. A. 2008, MNRAS 385, 206\\\
Dziembowski, W. A., Moskalik, P., Pamyatnykh, A. A. 1993, MNRAS 265, 588\\
Gautschy, A., Saio, H. 1993, MNRAS 262, 213\\
Iglesias, C.~A., Rogers, F.J. and Wilson, B. G. 1987, \apj {\bf 322}, L24\\
Iglesias, C. A., Rogers, F. J., Wilson, B. G. 1992, ApJ 397, 717\\
Iglesias C. A., Rogers F. J. 1996, ApJ, 464, 943\\
Kiriakidis, M., El Eid, M. F. and Glatzel, W. 1992, MNRAS 255, 1\\
Ledoux, P. 1951, ApJ, 114, 373\\
Lee U., Saio H. 1989, MNRAS, 237, 875\\
Miglio A., Montalb\'an J., Dupret M.-A., 2007, MNRAS, 37, L21\\
Pamyatnykh, A. A. 1999, Acta Astron., 49, 119\\
Moskalik, P., Dziembowski, W.~A. 1992, A\&A 256, L5\\
Pamyatnykh, A. A., Handler, G., Dziembowski, W. A. 2004, MNRAS 350, 1022\\\
Seaton, M. 1993,  ASP Conf. Ser., Vol. 40, p. 222\\
Seaton, M. 1999, MNRAS 307, 1008\\
Seaton, M. 2005, MNRAS 362, L1\\
Simon, N. R. 1982, ApJ  260, L87\\
Smolec, R., Moskalik, P. 2007, MNRAS, 277, 645\\
Stellingwer, R. F. 1978, AJ, 83, 1184\\
Townsend, R.H. D. 2005, MNRAS 360, 465\\
Waelkens, C. 1991, A\&A 246, 453\\}

\end{document}

%% file: CoAst_liege-proceedingslayo.tex
\renewcommand{\chaptermark}[1]{\markboth{#1}{}}

\newcommand{\apj}{ApJ}

\newcommand{\kopf}{\small\itshape Comm. in Asteroseismology \\ Contribution to the Proceedings of the 38$^{th}$\,LIAC\,/\,HELAS-ESTA\,/\,BAG, 2008
}

\newcommand{\Authors}[1]{\begin{center}\normalsize\bf\sf #1 \end{center}}

\renewcommand{\author}[1]{\begin{center}\normalsize\bf\sf #1 \end{center}}
\newcommand{\Address}[1]{\begin{center}\small\sf #1 \end{center}}

\DeclareGraphicsExtensions{.eps,.jpg}

\renewenvironment{abstract}{\section*{Abstract}\normalsize\sf}{}
\newcommand{\References}[1]{\begin{flushleft}{\large References\\}\vspace*{2mm}\small #1 \end{flushleft}}

\newcommand{\chapterCoAst}[2]{\chapter[\sf\normalsize #1\\ \footnotesize \hspace*{5mm}by #2 \sf\normalsize][]{#1\\}\rhead[\fancyplain{}{\sf\footnotesize \center{#1}}]{\fancyplain{}{\sffamily\thepage}}\lhead[\fancyplain{\kopf}{\sffamily\thepage}]{\fancyplain{\kopf}{\sf\footnotesize \center{#2}}}}

\newcommand{\figureCoAst}[5]{\begin{figure}[#4]
\centering
\includegraphics*[#5]{#1}
\caption{#2}
\label{#3}
\end{figure}}

\renewcommand{\chaptername}{}

\newcommand{\acknowledgments}[1]{\vspace*{5mm}\noindent  \textbf{Acknowledgments.} #1}